\documentstyle[prl,aps]{revtex}
\begin{document}


\vskip 4mm

\centerline{\large \bf Critical temperature of anisotropic superconductor}
\centerline{\large \bf containing both nonmagnetic and magnetic impurities}

\vskip 2mm

\centerline{Leonid A. Openov}

\vskip 2mm

\centerline{\it Moscow State Engineering Physics Institute
(Technical University)}
\centerline{\it 115409 Moscow, Russia}

\vskip 4mm

\begin{quotation}

The combined effect of both nonmagnetic and magnetic impurities on the
superconducting transition temperature is studied theoretically within the
BCS model. An expression for the critical temperature as a function of
potential and spin-flip scattering rates is derived for a two-dimensional
superconductor with arbitrary in-plane anisotropy of the superconducting
order parameter, ranging from isotropic $s$-wave to $d$-wave (or any pairing
state with nonzero angular momentum) and including anisotropic $s$-wave and
mixed $(d+s)$-wave as particular cases. This expression generalizes the
well-known Abrikosov-Gor'kov formula for the critical temperature of impure
superconductors. The effect of defects and impurities in high temperature
superconductors is discussed.

\end{quotation}

\vskip 6mm

\centerline{\bf 1. INTRODUCTION}

\vskip 2mm

The mechanism of high-$T_c$ superconductivity still remains unknown. It is
generally believed that elucidation of the symmetry of the superconducting
order parameter $\Delta({\bf p})$ in high-$T_c$ superconductors (HTSCs) could
narrow the list of pairing mechanisms debated. Since the CuO$_2$ layers are
thought to be responsible for the superconducting pairing in HTSCs, the
in-plane symmetry of $\Delta({\bf p})$ is of primary interest. However, in
spite of strong evidence provided for $d_{x^2-y^2}$ in-plane symmetry of
$\Delta({\bf p})$, experimental data are somewhat controversial
\cite{Wollman,Chaudhari,Buan1,Tsuei1,Kelley,Brawner,Van,Kleiner,Kendziora,%
Kane,Buan2,Ding,Bill,Leggett,Annett,Tsuei2} (though the recent research seems
to resolve the contradiction in favor of a $d_{x^2-y^2}$-wave
\cite{Kouznetsov}). The situation is complicated by orthorhombic distortion
of CuO$_2$ layers in some HTSCs, resulting in probable admixture of $s$-wave
component to otherwise pure $d$-wave $\Delta({\bf p})$ \cite{Sigrist,Emery}.
Since HTSCs differ structurally, the form of $\Delta({\bf p})$ may appear to
be material-dependent.

One indirect way to distinguish a pure $d$-wave from a highly anisotropic
$s$-wave or a mixed $(d+s)$-wave is to study the response of HTSCs to
intentionally incorporated impurities or radiation-induced defects. Depending
on the symmetry of $\Delta({\bf p})$, clear differences are predicted for the
variation of experimentally accessible characteristics such as the critical
temperature $T_c$, the superfluid density, {\it etc} \cite{Abrikosov1,Hotta,%
Borkowski1,Fehrenbacher,Hirashima,Y.Sun,Borkowski2,H.Kim,Beal-Monod,Preosti,%
Pokrovsky,Openov,Haran2}. For example, the rate of $T_c$ degradation by
defects and impurities in a two-dimensional superconductor is determined by
the value of the Fermi surface (FS) average
$\langle\Delta({\bf p})\rangle_{FS}$ \cite{Abrikosov1} and should be
different in $d$-wave superconductors with
$\langle\Delta({\bf p})\rangle_{FS}=0$ and anisotropic $s$-wave or mixed
$(d+s)$-wave ones with $\langle\Delta({\bf p})\rangle_{FS}\neq 0$, the
specific value of $\langle\Delta({\bf p})\rangle_{FS}$ being dictated by the
degree of $\Delta({\bf p})$ anisotropy or by the relative contributions of
$d$-wave and $s$-wave components to $\Delta({\bf p})$.

Numerous experimental studies give evidence for $T_c$ degradation by impurity
doping \cite{Cieplak,Chien,Zhao,Hedt,Castro,Prabhu,Lal,Walker,Kluge,Odagawa,%
Ilonca,Fukuzumi,Axnas,Quitmann,Brinkmann,Kuo} or radiation damage
\cite{Valles,Giapintzakis,Jackson,Rullier,Elesin,Weaver,Tolpygo,Lin,Krash1,%
Ginsberg} of HTSCs, the two effects being remarkably analogous if the
dependence of $T_c$ on the in-plane residual resistivity $\rho_0$ is
considered \cite{A.G.Sun}. Impurity-induced scattering of charge carriers in
doped HTSCs and their scattering by displaced host atoms in irradiated HTSCs
are believed to be the main reasons for the suppression of superconductivity
and the increase in $\rho_0$ \cite{Zhao,Hedt,Fukuzumi,Axnas,Jackson,Rullier,%
Weaver,Tolpygo}.

The comparison of experimental curves $T_c(\rho_0)$ with theoretical ones
reveals that the observed reduction of $T_c$ by impurities and radiation
defects is more gradual than predicted theoretically for $d$-wave
superconductors \cite{Openov,Giapintzakis,Elesin,Lin,Krash1,A.G.Sun,Radtke,%
Levin}. A critical value of $\rho_0^c$ at which $T_c=0$ ranges from 200
$\mu \Omega cm$ to 1500  $\mu \Omega cm$ depending on the type of disorder
and the kind of HTSC material \cite{Chien,Walker,Odagawa,Fukuzumi,Brinkmann,%
Rullier,Elesin,Tolpygo,Krash1,Ginsberg,A.G.Sun}, while for a $d$-wave
superconductor with $T_c \approx$ 100 K the theory gives
$\rho_0^c \approx 50~\mu \Omega cm$ \cite{Openov,A.G.Sun,Radtke,Levin}.
To reconcile the experimental findings with the $d$-wave symmetry of
$\Delta({\bf p})$ in HTSCs, a number of suggestions have been made, including
the anisotropy of impurity scattering in the momentum space \cite{Haran1,%
Kulic}, an "intermediate" (between Cooper pairs and local bosons) state of
paired electrons \cite{Sadovskii}, a depletion of the hole density due to the
oxygen vacancies in the CuO$_2$ planes \cite{Gupta}, an anomalously small
value of the plasma frequency \cite{Lin,Ginsberg}, the spatial variation of
the order parameter \cite{Franz}, {\it etc} \cite{Combescot}.

Another way is to abandon the $d$-wave hypothesis in favor of anisotropic
$s$-wave or mixed $(d+s)$-wave models \cite{H.Kim,Beal-Monod,Pokrovsky,%
Openov,Krash2} (it must be emphasized that if the relative weight of the
isotropic $s$-wave component in the mixed $(d+s)$-wave $\Delta({\bf p})$ is
large, then the symmetry of such an order parameter can in fact be viewed as
the anisotropic $s$-wave one \cite{Elesin}). However, while the initial slope
of experimentally observed $T_c(\rho_0)$ curve in HTSCs can actually be
explained by anisotropic $s$-wave symmetry of $\Delta({\bf p})$
\cite{Openov,Elesin,Krash1}, the theory faces problems when explaining the
complete suppression of superconductivity at a finite value of $\rho_0$.
Indeed, the theory predicts \cite{Abrikosov1,H.Kim,Pokrovsky,Openov} that
$T_c$ of a non-$d$-wave anisotropic two-dimensional superconductor doesn't
vanish at a certain critical value of $\rho_0^c$ (as it does in the case of
pure $d$-wave symmetry of $\Delta({\bf p})$), but instead asymptotically goes
to zero as $\rho_0$ increases. This contradicts the experiments mentioned
above. Besides, the experimentally observed form of the $T_c(\rho_0)$ curve
is usually close to linear \cite{Odagawa,Fukuzumi,Brinkmann,Rullier,Elesin,%
Tolpygo,Krash1,Ginsberg,A.G.Sun}, while the theory predicts a positive
curvature of the $T_c(\rho_0)$ curve in a non-$d$-wave superconductor.

Note, however, that experimentally determined values of $\rho_0$ reflect the
contribution from different scattering channels, while a theoretical analysis
of $T_c$ degradation by defects and impurities in HTSCs is usually made
for the specific case of spin-independent scattering potential
\cite{Abrikosov1,Fehrenbacher,Y.Sun,H.Kim,Beal-Monod,Preosti,Pokrovsky,%
Openov,Haran2,Elesin,Tolpygo,Krash1,Radtke,Levin,Haran1,Sadovskii,Gupta,%
Franz,Combescot}. Meanwhile a lot of experiments give evidence for the
presence of magnetic scatterers (along with nonmagnetic ones) in
non-stoichiometric HTSCs, e.g., in oxygen-deficient, doped or irradiated
samples \cite{Hedt,Axnas,Emerson,Togonidze,Xu,Finkel'stein,Xiao,Mahajan,%
Jayram,Mizuhashi,Williams,Awana}. For example, the oxygen vacancies or excess
oxygen atoms carry or induce local magnetic moments and hence play a role of
paramagnetic centers \cite{Emerson,Togonidze,Xu}. Furthermore, doping by Zn
induces local magnetic moments residing probably on the nearest-neighbor Cu
sites \cite{Finkel'stein,Xiao,Mahajan,Jayram,Mizuhashi,Williams,Awana}. This
is supported by studies of nonmagnetic impurities in Heisenberg
antiferromagnets \cite{Ng,Sandvik,Khaliullin} and by numerical calculations
within the two-dimensional $t-J$ model \cite{Odashima}. Besides, there are
intrinsic (host) magnetic atoms in some stoichiometric HTSCs, e.g., in
GdBa$_2$Cu$_3$O$_7$ \cite{Hor}.

Though the estimated moment-carrier exchange energy $J$ may appear to be too
small to solely account for suppression of superconductivity in disordered
HTSCs \cite{Walstedt}, an intriguing possibility of the magnetic
pair-breaking scattering as the common origin of the significant decrease in
$T_c$ remains \cite{Westerholt}. Moreover, it is suggested \cite{Kresin} that
interaction between Cooper pairs and localized magnetic moments in
"optimally" doped HTSCs leads to a depression of $T_c$ relative to its
"intrinsic" value. Hence, among other things, the understanding of the role
of magnetic scattering in HTSCs is important from the viewpoint of search for
new materials with higher $T_c$.

Since, first, there exist two channels of carrier scattering by magnetic
impurities (potential and spin-flip ones) and, second, in general both
magnetic and nonmagnetic scatterers are present in HTSCs, there is a need for
a theoretical model which could describe the effects of nonmagnetic and
magnetic scattering on equal footing. For an isotropic $s$-wave
superconductor this is the Abrikosov-Gor'kov theory \cite{Abrikosov2} which
predicts a rapid $T_c$ suppression by magnetic impurities and insensitivity
of $T_c$ to nonmagnetic scattering, in accordance with the Anderson theorem
\cite{Anderson} (a discussion about the validity of the Abrikosov-Gor'kov
approach \cite{Y.-J.Kim1,Abrikosov3} seems to be resolved in favor of the
standard Green's functions technique). However, the Abrikosov-Gor'kov formula
for $T_c$ versus scattering rate is not applicable to anisotropic
superconductors, no matter what the specific symmetry of $\Delta({\bf p})$ is
($d$-wave, $(d+s)$-wave, anisotropic $s$-wave or somewhat else). Hence, this
formula cannot be used to describe impurity effects in HTSCs. On the other
hand, theoretical considerations of impurity scattering in anisotropic
superconductors are commonly restricted to nonmagnetic scatterers only
\cite{Abrikosov1,Hotta,Fehrenbacher,Hirashima,Y.Sun,H.Kim,Beal-Monod,%
Preosti,Pokrovsky,Openov,Haran2,Radtke,Haran1,Sadovskii}. Such a status of
the theory of impure superconductors results in situations when the
experiments on the $T_c$ reduction by impurities or radiation-induced defects
in HTSCs are compared with {\it either} the Abrikosov-Gor'kov formula for
$T_c$ of {\it isotropic} $s$-wave superconductor containing {\it magnetic}
impurities \cite{Prabhu,Brinkmann,Jackson,Togonidze} {\it or} the formula for
$T_c$ of {\it anisotropic} superconductor but containing {\it nonmagnetic}
impurities only \cite{Kluge,Jackson,Elesin,Weaver,Tolpygo}. In the latter
case an {\it a priori} suggestion is often made about pure $d$-wave symmetry
of $\Delta({\bf p})$ \cite{Giapintzakis,Lin,Ginsberg}.

In a recent study \cite{Golubov}, the Abrikosov-Gor'kov theory has been
generalized to the case of a multiband superconductor with arbitrary
anisotropy of interband order parameter and arbitrary strength of magnetic
and/or nonmagnetic impurity scattering. Note, however, that the concept of
multiband superconductivity (arising, e.g., from CuO$_2$ plains and Cu-O
chains) is hardly probable to be applicable to HTSCs. Although it was argued
in \cite{Golubov} that the mathematical formalism had been proven to be the
same for a multiband superconductor and a superconductor with a general
angular anisotropy of the order parameter \cite{Allen}, an explicit formula
for the critical temperature of a one-band anisotropic superconductor
containing both nonmagnetic and magnetic impurities was not offered in
\cite{Golubov}.

The goal of this paper is to work out a theoretical framework for a
description of {\it combined} effect of nonmagnetic and magnetic scatterers
on $T_c$ of a two-dimensional superconductor with anisotropic $\Delta(\bf p)$
(preliminary results have been presented in \cite{Openov2}). We seek to
obtain a rather simple (free from needless theoretical complications)
Abrikosov-Gor'kov-like formula for $T_c$ which included physically meaningful
parameters and could be compared with available experimental data. Within the
weak coupling limit of the BCS model and without specifying the microscopic
mechanism of superconducting pairing, we derive the expression that relates
$T_c$ to relaxation rates of charge carriers by nonmagnetic and magnetic
scatterers, as well as to the numerical coefficient $\chi=%
1-\langle\Delta({\bf p})\rangle^2_{FS}/\langle\Delta^2({\bf p})\rangle_{FS}$
which is a measure of the degree of in-plane anisotropy of $\Delta({\bf p})$
on the FS. The range $0\leq\chi\leq 1$ covers the cases of isotropic $s$-wave
($\chi=0$), $d$-wave ($\chi=1$), anisotropic $s$-wave ($0<\chi<1$), and mixed
$(d+s)$-wave ($0<\chi<1$) symmetries of $\Delta({\bf p})$. In two particular
cases of ($i$) both nonmagnetic and magnetic scattering in an isotropic
$s$-wave superconductor ($\chi =0$) and ($ii$) nonmagnetic scattering only in
a superconductor with arbitrary anisotropy of $\Delta({\bf p})$
($0\leq\chi\leq 1$), our expression for $T_c$ reduces to the well-known
formulae \cite{Abrikosov1,Abrikosov2}.

The paper is organized as follows. The BCS model for an impure anisotropic
superconductor containing both nonmagnetic and magnetic scatterers is
described in Sec.2 along with the theoretical formalism. The expression for
the critical temperature as a function of potential and spin-flip relaxation
times of charge carriers is derived in Sec.3 for a superconductor with an
arbitrary degree of the order parameter anisotropy. The results obtained are
discussed in Sec.4. In Sec.5 concluding remarks are given.

\vskip 6mm

\centerline{\bf 2. MODEL AND FORMALISM}

\vskip 2mm

Within the framework of the BCS model, the Hamiltonian of a superconductor
containing both nonmagnetic and magnetic impurities is as follows
\begin{equation}
\hat{H}=\sum_{{\bf p},\sigma}\xi ({\bf p})\hat{a}^{+}_{{\bf p}\sigma}%
\hat{a}^{}_{{\bf p}\sigma}~+\sum_{{\bf p},{\bf p}^\prime,\sigma,\sigma^%
\prime}U({\bf p},\sigma;{\bf p}^\prime,\sigma^\prime)\hat{a}^{+}_{{\bf p}%
\sigma}\hat{a}^{}_{{\bf p}^\prime\sigma^\prime}~+\sum_{{\bf p},{\bf p}^%
\prime}V({\bf p},{\bf p}^\prime)\hat{a}^{+}_{{\bf p}\uparrow}\hat{a}^{+}_%
{-{\bf p}\downarrow}\hat{a}^{}_{-{\bf p}^\prime\downarrow}\hat{a}^{}_%
{{\bf p}^\prime\uparrow},
\label{Hamiltonian}
\end{equation}
where the operator $\hat{a}^{+}_{{\bf p}\sigma} (\hat{a}^{}_{{\bf p}\sigma})$
creates (annihilates) an electron with the quasimomentum ${\bf p}$ and the
spin projection on $z$-axis $\sigma$ = $\uparrow$ or $\downarrow$,
$\xi({\bf p})=\epsilon({\bf p})-\mu$ is the (spin independent) quasiparticle
energy measured from the chemical potential $\mu$,
$U({\bf p},\sigma; {\bf p}^\prime,\sigma^\prime)$ is the matrix element for
electron scattering by randomly distributed impurities (defects) from the
state $({\bf p}^\prime,\sigma^\prime)$ to the state $({\bf p},\sigma)$, and
$V({\bf p},{\bf p}^\prime)$ is the BCS pair potential.

Let the following sum
\begin{equation}
U({\bf r})=U_n({\bf r})+U_m({\bf r})
\label{Interaction}
\end{equation}
be the total interaction between a conduction electron at a point ${\bf r}$
and {\it all} impurities present in the sample, $U_n({\bf r})$ and
$U_m({\bf r})$ being the interaction components due to nonmagnetic and
magnetic impurities respectively:
\begin{equation}
U_n({\bf r})=\sum_{\alpha}u_n({\bf r} - {\bf R}_{\alpha})~,~~~
U_m({\bf r})=\sum_{\beta}u_m({\bf r} - {\bf R}_{\beta})~,
\label{Int_n_m}
\end{equation}
where $u_n({\bf r} - {\bf R}_{\alpha})$ is the interaction between an
electron at ${\bf r}$ and a nonmagnetic impurity at ${\bf R}_{\alpha}$, while
$u_m({\bf r} - {\bf R}_{\beta})$ is the interaction between an electron at
${\bf r}$ and a magnetic impurity at ${\bf R}_{\beta}$. Since, in general,
magnetic impurities give rise to both potential and exchange scattering
\cite{Abrikosov2}, one has
\begin{equation}
u_m({\bf r} - {\bf R}_{\beta}) = u_m^{pot}({\bf r} - {\bf R}_{\beta}) + %
u_m^{ex}({\bf r} - {\bf R}_{\beta}),
\label{Int_m}
\end{equation}
where $u_m^{pot}({\bf r} - {\bf R}_{\beta})$ is the spin-independent
potential component, and
\begin{equation}
u_m^{ex}({\bf r} - {\bf R}_{\beta}) = %
J({\bf r} - {\bf R}_{\beta}){\bf s}{\bf S}_{\beta}
\label{Exchange}
\end{equation}
is the exchange interaction. Here $J({\bf r} - {\bf R}_{\beta})$ is the
exchange energy, ${\bf S}_{\beta}$ is the spin of magnetic impurity located
at ${\bf R}_{\beta}$ and ${\bf s}={\bf \sigma}/2$ is the operator of electron
spin (the three components of ${\bf \sigma}$ are the Pauli matrices
$\sigma_x$, $\sigma_y$, $\sigma_z$). We shall assume that orientations of
the paramagnetic spins ${\bf S}_{\beta}$ are fixed and remain unchanged upon
electron scattering (taking into account dynamic transitions of the impurity
spin between $2S+1$ magnetic sublevels has a little effect on the results
\cite{Y.-J.Kim2}).

The electron interactions $u_n({\bf r} - {\bf R}_{\alpha})$ and
$u_m^{pot}({\bf r} - {\bf R}_{\beta})$ with nonmagnetic and magnetic
impurities respectively are spin-independent and hence contribute to the
scattering matrix elements $U({\bf p},\sigma; {\bf p}^\prime,\sigma)$ only.
In its turn, exchange interactions of electron with magnetic impurities,
$u_m^{ex}({\bf r} - {\bf R}_{\beta})$, result in both spin-conserving and
spin-flip scattering events and hence contribute to
$U({\bf p},\sigma; {\bf p}^\prime,\sigma)$ as well as to
$U({\bf p},\sigma; {\bf p}^\prime,-\sigma)$. Let us write down the matrix
element $U({\bf p},\sigma; {\bf p}^\prime,\sigma^\prime)$ as
\begin{equation}
U({\bf p},\sigma;{\bf p}^\prime,\sigma^\prime)~=~
U_1({\bf p},{\bf p}^\prime,\sigma)\delta_{\sigma,\sigma^\prime}~+~
U_2({\bf p},{\bf p}^\prime,\sigma)\delta_{\sigma,-\sigma^\prime}~,
\label{mat_el}
\end{equation}
where
\begin{eqnarray}
&&U_1({\bf p},{\bf p}^\prime,\sigma)~=~u_n({\bf p},{\bf p}^\prime)%
\sum_{\alpha}e^{-i({\bf p}-{\bf p}^\prime){\bf R}_{\alpha}}~+~%
u_m^{pot}({\bf p},{\bf p}^\prime)%
\sum_{\beta}e^{-i({\bf p}-{\bf p}^\prime){\bf R}_{\beta}}~+~%
\frac{1}{2}J({\bf p},{\bf p}^\prime)\gamma_{\sigma}%
\sum_{\beta}e^{-i({\bf p}-{\bf p}^\prime){\bf R}_{\beta}}S_{\beta}^z~,
\nonumber \\
&&U_2({\bf p},{\bf p}^\prime,\sigma)~=~\frac{1}{2}J({\bf p},{\bf p}^\prime)%
\sum_{\beta}e^{-i({\bf p}-{\bf p}^\prime){\bf R}_{\beta}}%
(S_{\beta}^x -i\gamma_{\sigma}S_{\beta}^y)~.
\label{U1,U2}
\end{eqnarray}
Here $u_n({\bf p},{\bf p}^\prime)$, $u_m^{pot}({\bf p},{\bf p}^\prime)$, and
$J({\bf p},{\bf p}^\prime)$ are the components of the matrix element for
electron scattering by an isolated impurity; $\gamma_{\sigma}$ = +1 and -1
for $\sigma$ = $\uparrow$ and $\downarrow$ respectively. We do not consider
the direct effect of nonmagnetic disorder on the magnetic pair breaking (this
effect has been studied in Refs. \cite{Y.-J.Kim2,Devereaux} for the case of
an isotropic $s$-wave pairing).

In order to account for anisotropy of the superconducting state, we assume a
factorizable phenomenological pairing interaction
$V({\bf p},{\bf p}^{\prime})$ of the form (see, e. g., \cite{Abrikosov1})
\begin{equation}
V({\bf p},{\bf p}^{\prime})=-V_0\phi({\bf n})\phi({\bf n}^{\prime}),
\label{V(p,p')}
\end{equation}
where $V_0$ is the pairing energy, ${\bf n}={{\bf p}}/p$ is a unit vector
along the momentum. Then the order parameter $\Delta ({\bf p})$ is
\cite{Abrikosov1}
\begin{equation}
\Delta({\bf p})=-\sum_{{\bf p}^\prime}V({\bf p},{\bf p}^\prime)\langle%
\hat{a}^{}_{-{\bf p}^\prime\downarrow}\hat{a}^{}_{{\bf p}^\prime\uparrow}%
\rangle=\Delta_0\phi({\bf n}),
\label{Delta}
\end{equation}
where $\Delta_0$ depends on the temperature. The function $\phi({\bf n})$
specifies the anisotropy of $\Delta ({\bf p})$ in the momentum space
(e.g., $\phi({\bf n})\equiv 1$ for isotropic $s$-wave pairing). We assume
that $\phi({\bf n})$ is temperature independent.

The self-consistent equation for $\Delta ({\bf p})$ can be derived by means
of Green's functions technique (see, e.g., \cite {Abrikosov4}). We define the
normal and anomalous temperature Green's functions
\begin{eqnarray}
&&G({\bf p},\sigma; {\bf p}^{\prime},\sigma^{\prime};\tau)~=~%
-\left<T_{\tau}\hat{a}^{}_{{\bf p}\sigma}(\tau)%
\hat{a}^{+}_{{\bf p}^{\prime}\sigma^{\prime}}(0)\right>, \nonumber \\
&&F({\bf p},\sigma; {\bf p}^{\prime},\sigma^{\prime};\tau)~=~%
\left<T_{\tau}\hat{a}^{+}_{-{\bf p}-\sigma}(\tau)%
\hat{a}^{+}_{{\bf p}^{\prime}\sigma^{\prime}}(0)\right>, \nonumber \\
&&\tilde{G}({\bf p},\sigma; {\bf p}^{\prime},\sigma^{\prime};\tau)~=~%
-\left<T_{\tau}\hat{a}^{+}_{-{\bf p}-\sigma}(\tau)%
\hat{a}^{}_{-{\bf p}^{\prime}-\sigma^{\prime}}(0)\right>, \nonumber \\
&&\tilde{F}({\bf p},\sigma; {\bf p}^{\prime},\sigma^{\prime};\tau)~=~%
\left<T_{\tau}\hat{a}^{}_{{\bf p}\sigma}(\tau)%
\hat{a}^{}_{-{\bf p}^{\prime}-\sigma^{\prime}}(0)\right>,
\label{G,F}
\end{eqnarray}
and their Fourier transforms
$G({\bf p},\sigma; {\bf p}^{\prime},\sigma^{\prime}; \omega)$,
$F({\bf p},\sigma; {\bf p}^{\prime},\sigma^{\prime}; \omega)$,
$\tilde{G}({\bf p},\sigma; {\bf p}^{\prime},\sigma^{\prime}; \omega)$,
$\tilde{F}({\bf p},\sigma; {\bf p}^{\prime},\sigma^{\prime}; \omega)$,
where angular brackets stand for the statistical averaging with the
Hamiltonian (\ref{Hamiltonian}), the symbol $T_\tau$ denotes the time
ordering, $\tau$ is the imaginary time, and $\omega=\pi T(2n+1)$ are
Matsubara frequencies (we set $\hbar=k_B=1$ throughout the paper).

It is convenient to introduce the matrix Green function
$\hat{G}({\bf p},\sigma; {\bf p}^{\prime},\sigma^{\prime};\omega)$
in the Nambu representation:
\begin{eqnarray}
\hat{G}({\bf p},\sigma; {\bf p}^{\prime},\sigma^{\prime};\omega)~=~
\left(
\begin{array}{cccc}
G({\bf p},\sigma; {\bf p}^{\prime},\sigma^{\prime};\omega) &\
-\tilde{F}({\bf p},\sigma; {\bf p}^{\prime},\sigma^{\prime};\omega) \\
-F({\bf p},\sigma; {\bf p}^{\prime},\sigma^{\prime};\omega) &\
\tilde{G}({\bf p},\sigma; {\bf p}^{\prime},\sigma^{\prime};\omega) \\
\end{array}
\right).
\label{matrG}
\end{eqnarray}
We stress that
$\hat{G}({\bf p},\sigma; {\bf p}^{\prime},\sigma^{\prime};\omega)$
is nondiagonal in spin space (since there is spin-flip scattering of
electrons on magnetic impurities) as well as in momentum space (until
averaged over impurities coordinates). The matrix equation for
$\hat{G}({\bf p},\sigma; {\bf p}^{\prime},\sigma^{\prime}; \omega)$
can be written as
\begin{equation}
\hat{G}_0^{-1}({\bf p},\sigma;{\bf k},\lambda;\omega)
\hat{G}({\bf k},\lambda; {\bf p}^{\prime},\sigma^{\prime};\omega)-
\hat{U}({\bf p},\sigma;{\bf k},\lambda)
\hat{G}({\bf k},\lambda; {\bf p}^{\prime},\sigma^{\prime};\omega)=
\hat{1}\delta_{{\bf p},{\bf p}^{\prime}}\delta_{\sigma,\sigma^\prime},
\label{mat-eq}
\end{equation}
where $\hat{G}_0({\bf p},\sigma;{\bf k},\lambda;\omega)$=%
$\hat{G}_0({\bf p},\sigma,\omega)%
\delta_{{\bf p},{\bf k}}\delta_{\sigma,\lambda}$
is the Green function of a clean sample,
\begin{equation}
\hat{G}_0({\bf p},\sigma,\omega)=-\frac{1}%
{\omega^2+\xi({\bf p})^2+|\Delta({\bf p})|^2}
\left(
\begin{array}{cccc}
i\omega+\xi({\bf p}) &\ \gamma_{\sigma}\Delta({\bf p}) \\
\gamma_{\sigma}\Delta^*({\bf p}) &\ i\omega-\xi({\bf p}) \\
\end{array}
\right),
\label{G_0}
\end{equation}
$\hat{1}$ is the unit matrix 2x2, and the matrix
$\hat{U}({\bf p},\sigma;{\bf k},\lambda)$ that describes the effect of
impurity scattering has the form
\begin{equation}
\hat{U}({\bf p},\sigma;{\bf k},\lambda)=
\left(
\begin{array}{cccc}
U({\bf p},\sigma;{\bf k},\lambda) &\ 0 \\
0 &\ -U({\bf p},-\lambda;{\bf k},-\sigma) \\
\end{array}
\right).
\label{U}
\end{equation}
The summation over repeated indices in Eq. (\ref{mat-eq}) and below is
implied. We note that $\hat{G}_0({\bf p},\sigma,\omega)$ depends on $\sigma$
through $\gamma_{\sigma}$.

In order to avoid needless mathematical complications and to express the
final results in terms of as few parameters as possible, we make several
simplifying assumptions: (i) we consider the short-range scattering
potentials, so that the matrix elements $u_n({\bf p},{\bf p}^\prime)$,
$u_m^{pot}({\bf p},{\bf p}^\prime)$, $J({\bf p},{\bf p}^\prime)$ are
momentum-independent and equal to $u_n$, $u_m^{pot}$, $J$ respectively
($s$-wave impurity scattering); (ii) we treat the impurity scattering in the
Born limit; (iii) we restrict the momenta of the electron self-energy and BCS
pair potential to the FS.

After averaging the Eq. (\ref{mat-eq}) over impurity configurations and
directions of impurity spins one has
\begin{equation}
\left<\hat{G}({\bf p},\sigma;{\bf p}^{\prime},\sigma^{\prime};\omega)\right>%
_{imp}=~\hat{G}({\bf p},\sigma,\omega)%
\delta_{{\bf p},{\bf p}^{\prime}}\delta_{\sigma,\sigma^{\prime}},
\label{G_average}
\end{equation}
where
\begin{equation}
\hat{G}^{-1}({\bf p},\sigma,\omega)=\hat{G}_0^{-1}({\bf p},\sigma,\omega)-%
\hat{M}({\bf p},\sigma,\omega),
\label{G^-1}
\end{equation}
\begin{eqnarray}
&&\hat{M}({\bf p},\sigma,\omega)=%
\left<\hat{U}({\bf p},\sigma;{\bf k},\lambda)\hat{G}({\bf k},\lambda,\omega)%
\hat{U}({\bf k},\lambda;{\bf p},\sigma)\right>_{imp}= \nonumber \\
&&\left(
\begin{array}{cccc}
\left(c_n|u_n|^2+c_m|u_m^{pot}|^2+c_m|u_m^{ex}|^2\right)%
\sum_{{\bf k}}G({\bf k},\sigma,\omega) &\
\left(c_n|u_n|^2+c_m|u_m^{pot}|^2-c_m|u_m^{ex}|^2\right)%
\sum_{{\bf k}}\tilde{F}({\bf k},\sigma,\omega) \\
\left(c_n|u_n|^2+c_m|u_m^{pot}|^2-c_m|u_m^{ex}|^2\right)%
\sum_{{\bf k}}F({\bf k},\sigma,\omega) &\
\left(c_n|u_n|^2+c_m|u_m^{pot}|^2+c_m|u_m^{ex}|^2\right)%
\sum_{{\bf k}}\tilde{G}({\bf k},\sigma,\omega) \\
\end{array}
\right).
\label{M}
\end{eqnarray}
Here $c_n$ and $c_m$ are the concentrations of nonmagnetic and magnetic
impurities respectively, and we have designated $|u_m^{ex}|^2=|J|^2S(S+1)/4$.
Note that $|u_m^{ex}|^2$ includes contributions from both spin-flip and
spin-conserving scattering of electrons due to their exchange interaction
with magnetic impurities, Eq. (\ref{Exchange}). This is because the matrix
element of the spin-conserving exchange scattering depends on the orientation
of electron spin through $\gamma_{\sigma}$, see Eq. (\ref{U1,U2}). The
coefficient $c_m|u_m^{ex}|^2$ in Eq. (\ref{M}) enters into the factors at the
normal and anomalous Green functions with opposite signs, while the
coefficients $c_n|u_n|^2$ and $c_m|u_m^{pot}|^2$, which are due to the
scattering by nonmagnetic impurities and the potential component of the
scattering by magnetic impurities respectively, appear in Eq. (\ref{M}) with
the same signs.

Making use of Eqs. (\ref{G_0}), (\ref{G^-1}), and (\ref{M}), we obtain
\begin{equation}
\hat{G}({\bf p},\sigma,\omega,\Delta)=%
\hat{G}_0({\bf p},\sigma,\omega^{\prime},\Delta_{\omega}),
\label{G}
\end{equation}
where
\begin{equation}
\omega^{\prime}=\omega-i(c_n|u_n|^2+c_m|u_m^{pot}|^2+c_m|u_m^{ex}|^2)%
\sum_{{\bf k}}\frac{i\omega^{\prime}+\xi({\bf k})}%
{\omega^{\prime 2}+\xi^2({\bf k})+|\Delta_\omega({\bf k})|^2}~,
\label{omega'}
\end{equation}
\begin{equation}
\Delta_\omega({\bf p})=\Delta({\bf p})+%
(c_n|u_n|^2+c_m|u_m^{pot}|^2-c_m|u_m^{ex}|^2)\sum_{{\bf k}}%
\frac{\Delta_\omega({\bf k})}{\omega^{\prime 2}+\xi^2({\bf k})+%
|\Delta_\omega({\bf k})|^2}~.
\label{Delta_omega}
\end{equation}
Then one has from Eq. (\ref{Delta}):
\begin{equation}
\Delta({\bf p})=-T\sum_{\omega}\sum_{{\bf p}^\prime}%
V({\bf p},{\bf p}^\prime)\frac{\Delta_\omega({\bf p}^\prime)}%
{\omega^{\prime 2}+\xi^2({\bf p}^\prime)+|\Delta_\omega({\bf p}^\prime)|^2}~.
\label{Delta1}
\end{equation}

For further considerations it is convenient to express the coefficients
$c_n|u_n|^2$, $c_m|u_m^{pot}|^2$, and $c_m|u_m^{ex}|^2$ in terms of electron
relaxation times $\tau_n$, $\tau_m^{pot}$, and $\tau_m^{ex}$ for scattering
by nonmagnetic impurities, potential scattering by magnetic impurities, and
exchange scattering by magnetic impurities respectively:
\begin{equation}
\frac{1}{\tau_n}=2\pi c_n|u_n|^2N(0),~%
\frac{1}{\tau_m^{pot}}=2\pi c_m|u_m^{pot}|^2N(0),~%
\frac{1}{\tau_m^{ex}}=2\pi c_m|u_m^{ex}|^2N(0),
\label{tau_nm}
\end{equation}
where $N(0)$ is the density of electron states at the Fermi level. The
electron relaxation time $\tau_m$ due to magnetic impurities is given by the
expression
\begin{equation}
\frac{1}{\tau_m}=\frac{1}{\tau_m^{pot}}+\frac{1}{\tau_m^{ex}},
\label{tau_m}
\end{equation}
while the total electron relaxation time $\tau$ due to all impurities present
in the sample can be found as
\begin{equation}
\frac{1}{\tau}=\frac{1}{\tau_n}+\frac{1}{\tau_m}=%
\frac{1}{\tau_n}+\frac{1}{\tau_m^{pot}}+\frac{1}{\tau_m^{ex}}.
\label{tau}
\end{equation}
We note that Eqs. (\ref{tau_nm}) allow one to express the final results in
terms of three relaxation times ($\tau_n$, $\tau_m^{pot}$, $\tau_m^{ex}$)
instead of a large number of unknown parameters such as impurity
concentrations and scattering matrix elements. Besides, the relaxation times
are associated with the residual resistivity. This facilitates a comparison
between the theory and experiment.

\vskip 6mm

\centerline{\bf 3. CRITICAL TEMPERATURE OF IMPURE ANISOTROPIC SUPERCONDUCTOR}

\vskip 2mm

The critical temperature $T_c$ can be found from Eqs. (\ref{omega'}),
(\ref{Delta_omega}), and (\ref{Delta1}) as the temperature at which the order
parameter goes to zero, i.e., $\Delta_0\rightarrow 0$ in Eq. (\ref{Delta}).
Setting $|\Delta_\omega({\bf k})|^2=0$ in the denominators of Eqs.
(\ref{omega'}) and (\ref{Delta_omega}) and taking Eqs. (\ref{tau_nm}) into
account, we have at $T\rightarrow T_c$:
\begin{equation}
\omega^\prime=\omega+\frac{1}{2}\left(1/\tau_n+1/\tau_m^{pot}+1/\tau_m^{ex}%
\right)sign(\omega),
\label{omega'2}
\end{equation}
\begin{equation}
\Delta_\omega({\bf p})=\Delta({\bf p})+\frac{1}{2|\omega^\prime|}\left(%
(1/\tau_n+1/\tau_m^{pot}-1/\tau_m^{ex}\right)%
\langle\Delta_\omega({\bf p})\rangle_{FS},
\label{Delta_omega2}
\end{equation}
where the angular brackets $\langle ... \rangle_{FS}$ stand for a FS average:
\begin{equation}
\langle ... \rangle_{FS}=\int_{FS}(...)\frac{d\Omega_{\bf p}}{|\partial%
\xi({\bf p})/\partial {\bf p}|}\Biggl/\int_{FS}\frac{d\Omega_%
{\bf p}}{|\partial\xi({\bf p})/\partial {\bf p}|}.
\label{average}
\end{equation}

Substituting Eqs. (\ref{omega'2}) and (\ref{Delta_omega2}) in
Eq. (\ref{Delta1}), setting $|\Delta_\omega({\bf p}^{\prime})|^2=0$ in the
denominator of Eq. (\ref{Delta1}), and taking Eqs. (\ref{V(p,p')}) and
(\ref{Delta}) into account, we have
\begin{equation}
\frac{1}{\lambda}=\pi T_c\sum_{\omega}\frac{1}{|\omega|+%
\frac{1}{2}\left(1/\tau_n+1/\tau_m^{pot}+1/\tau_m^{ex}\right)}%
\left[\langle\phi^2({\bf n})\rangle_{FS}+\langle\phi({\bf n})\rangle^2_{FS}%
\frac{1/\tau_n+1/\tau_m^{pot}-1/\tau_m^{ex}}%
{2\left(|\omega|+1/\tau_m^{ex}\right)}\right],
\label{Tc}
\end{equation}
where $\lambda=V_0N(0)$ is the electron-boson coupling constant. The equation
for the critical temperature $T_{c0}$ in the absence of impurities (i.e., at
$1/\tau_n=1/\tau_m^{pot}=1/\tau_m^{ex}=0$) reads
\begin{equation}
\frac{1}{\lambda}=\pi T_{c0}\langle\phi^2({\bf n})\rangle_{FS}%
\sum_{\omega}\frac{1}{|\omega|}.
\label{Tc0}
\end{equation}
Following the standard procedure, we obtain from Eqs. (\ref{Tc}) and
(\ref{Tc0}) the equation for the critical temperature $T_c$ as
\begin{equation}
\ln\left(\frac{T_{c0}}{T_c}\right)=\pi T_c\sum_{\omega}\frac{1}{|\omega|+%
\frac{1}{2}\left(1/\tau_n+1/\tau_m^{pot}+1/\tau_m^{ex}\right)}\left[%
\frac{1}{2|\omega|}\left(1/\tau_n+1/\tau_m^{pot}+1/\tau_m^{ex}\right)-%
\frac{\langle\phi({\bf n})\rangle^2_{FS}}{\langle\phi^2({\bf n})\rangle_{FS}}%
\frac{1/\tau_n+1/\tau_m^{pot}-1/\tau_m^{ex}}{2\left%
(|\omega|+1/\tau_m^{ex}\right)}\right].
\label{ln(Tc0/Tc)}
\end{equation}

At this stage it is convenient to introduce the coefficient $\chi$ of
anisotropy of the order parameter on the FS \cite{Abrikosov1,Openov}
\begin{equation}
\chi=1-\frac{\langle\phi({\bf n})\rangle^2_{FS}}%
{\langle\phi^2({\bf n})\rangle_{FS}}=%
1-\frac{\langle\Delta({\bf p})\rangle^2_{FS}}%
{\langle\Delta^2({\bf p})\rangle_{FS}}.
\label{chi}
\end{equation}
For isotropic $s$-wave pairing we have $\Delta({\bf p})\equiv const$ on the
FS; therefore, $\langle\Delta({\bf p})\rangle^2_{FS}%
=\langle\Delta^2({\bf p})\rangle_{FS}$, and $\chi=0$. For a two-dimensional
superconductor with $d$-wave pairing we have $\chi=1$ since
$\langle\Delta({\bf p})\rangle_{FS}=0$. The range $0<\chi<1$ corresponds to
anisotropic $s$-wave or mixed $(d+s)$-wave in-plane pairing. The higher the
in-plane anisotropy of $\Delta({\bf p})$ (e.g., the greater the partial
weight of a $d$-wave in the case of mixed pairing), the closer to unity is
the value of $\chi$.

Note that $\chi=1$ holds not only for $d$-wave pairing state, but also for
any pairing state with angular momentum $l>0$, e.g. for $p$-wave state
($l=1$), see Eq. (\ref {chi}). In its turn, the range $0<\chi<1$ generally
corresponds to mixing of $s$-wave state with some higher angular harmonic
state. Hence, while this paper focuses primarily on $s$-wave, $d$-wave, and
$(d+s)$-wave states, one should keep in mind that the results obtained are
more general and may be applied to superconductors with other symmetries of
the order parameter.

Making use of the definition (\ref{chi}) and the formula \cite{Abramowitz}
\begin{equation}
\sum_{k=0}^{\infty}\left(\frac{1}{k+x}-\frac{1}{k+y}\right)=\Psi(y)-\Psi(x),
\label{digamma}
\end{equation}
where $\Psi$ is the digamma function, we obtain from Eq. (\ref{ln(Tc0/Tc)}):
\begin{equation}
\ln\left(\frac{T_{c0}}{T_c}\right)=(1-\chi)\left[\Psi\left(\frac{1}{2}+%
\frac{1}{2\pi T_c\tau_m^{ex}}\right)-\Psi\left(\frac{1}{2}\right)\right]+%
\chi\left[\Psi\left(\frac{1}{2}+\frac{1}{4\pi T_c}\left(\frac{1}{\tau_n}+%
\frac{1}{\tau_m^{pot}}+\frac{1}{\tau_m^{ex}}\right)\right)-%
\Psi\left(\frac{1}{2}\right)\right].
\label{ln(Tc0/Tc)2}
\end{equation}
In two particular cases of ($i$) both nonmagnetic and magnetic scattering in
an isotropic $s$-wave superconductor ($\chi=0$) and ($ii$) nonmagnetic
scattering only in a superconductor with arbitrary in-plane anisotropy of
$\Delta({\bf p})$ ($1/\tau_m^{ex}=1/\tau_m^{pot}=0$, $0\leq\chi\leq 1$), the
Eq. (\ref{ln(Tc0/Tc)2}) reduces to well-known expressions \cite{Abrikosov2,%
Abrikosov1}
\begin{equation}
\ln\left(\frac{T_{c0}}{T_c}\right)=\Psi\left(\frac{1}{2}+%
\frac{1}{2\pi T_c\tau_m^{ex}}\right)-\Psi\left(\frac{1}{2}\right)
\label{ln(Tc0/Tc)3}
\end{equation}
and
\begin{equation}
\ln\left(\frac{T_{c0}}{T_c}\right)=\chi\left[\Psi\left(\frac{1}{2}+%
\frac{1}{4\pi T_c\tau_n}\right)-\Psi\left(\frac{1}{2}\right)\right].
\label{ln(Tc0/Tc)4}
\end{equation}
respectively.

Now let us consider the limiting cases of low and high impurity concentration
($T_{c0}-T_c<<T_{c0}$ and $T_c\rightarrow 0$ respectively). At
$1/4\pi T_{c0}\tau_n<<1$, $1/4\pi T_{c0}\tau_m^{pot}<<1$ and
$1/4\pi T_{c0}\tau_m^{ex}<<1$ (low impurity concentration) one has from
Eq. (\ref{ln(Tc0/Tc)2}):
\begin{equation}
T_{c0}-T_c\approx\frac{\pi}{4}\left[\frac{\chi}{2}\left(\frac{1}{\tau_n}+%
\frac{1}{\tau_m^{pot}}\right)+\frac{1-\chi/2}{\tau_m^{ex}}\right].
\label{Tc1}
\end{equation}
In particular cases ($i$) and ($ii$) considered above, Eq. (\ref{Tc1})
reduces to expressions \cite{Abrikosov2,Abrikosov1}
\begin{equation}
T_{c0}-T_c\approx\frac{\pi}{4\tau_m^{ex}}
\label{Tc2}
\end{equation}
and
\begin{equation}
T_{c0}-T_c\approx\frac{\pi\chi}{8\tau_n}
\label{Tc3}
\end{equation}
for initial $T_c$ suppression by magnetic (at $\chi=0$) or nonmagnetic (at
arbitrary value of $\chi$) scatterers respectively.

As to the high impurity concentration, we recall that in the BCS theory,
nonmagnetic scattering alone is insufficient for the non-$d$-wave
two-dimensional superconductivity ($0\leq\chi<1$) to be destroyed completely
\cite{Abrikosov1}; at $1/\tau_m^{ex}=0$, the value of $T_c$ asymptotically
goes to zero as $1/\tau_n$ increases. On the other hand, $T_c$ of a $d$-wave
superconductor with $\chi=1$ vanishes at a critical value
$1/\tau_{n,c}=\pi T_{c0}/\gamma\approx 1.764T_{c0}$, with
$\gamma=e^C\approx 1.781$, where $C$ is the Euler constant. In its turn,
magnetic scattering in the absence of nonmagnetic scattering ($1/\tau_n=0$)
is known to suppress the isotropic $s$-wave superconductivity with $\chi=0$
at a critical value $1/\tau_{m,c}^{ex}=\pi T_{c0}/2\gamma\approx 0.882T_{c0}$
(Ref. \cite{Abrikosov2}).

On the basis of Eq. (\ref{ln(Tc0/Tc)2}), it is straightforward to derive the
general condition for impurity (defect) suppression of $T_c$ for a
superconductor having an arbitrary in-plane anisotropy coefficient $\chi$ and
containing both nonmagnetic and magnetic scatterers:
\begin{equation}
\frac{1}{\tau_{eff,c}}=\frac{\pi}{\gamma}2^{\chi-1}T_{c0},
\label{tau_eff,c}
\end{equation}
where $\tau_{eff,c}$ is the critical value of the effective relaxation time
$\tau_{eff}$, defined as
\begin{equation}
\frac{1}{\tau_{eff}}=\left(\frac{1}{\tau_m^{ex}}\right)^{1-\chi}%
\left(\frac{1}{\tau_n}+\frac{1}{\tau_m^{pot}}+%
\frac{1}{\tau_m^{ex}}\right)^{\chi}.
\label{tau_eff}
\end{equation}

From Eqs. (\ref{tau_eff,c}) and (\ref{tau_eff}) one can see that
$1/\tau_{eff,c}$ increases monotonically with $1/\tau_n$, $1/\tau_m^{pot}$,
and $1/\tau_m^{ex}$ at any value of $\chi$, with the exception of the case
$\chi=0$, where $1/\tau_{eff,c}$ doesn't depend on $1/\tau_n$ and
$1/\tau_m^{pot}$, see Eq. (\ref{tau_eff}). If $\chi$ is close to unity
($\Delta(\bf p)$ with strong in-plane anisotropy), then
$1/\tau_{eff}\approx 1/\tau_n+1/\tau_m^{pot}+1/\tau_m^{ex}$, i.e., the
contribution of nonmagnetic and magnetic scattering to pair breaking is about
the same. If $\chi<<1$ (almost isotropic $\Delta(\bf p)$), then
$1/\tau_{eff}\approx 1/\tau_m^{ex}$, i.e., $\tau_{eff}$ is determined
primarily by magnetic scattering. The higher the anisotropy coefficient
$\chi$, the greater is the relative contribution of nonmagnetic scatterers
to $T_c$ suppression as compared to magnetic scatterers.

We note however that while the concept of the effective relaxation time
$\tau_{eff}$ can be used for evaluation of the {\it critical level} of
nonmagnetic and magnetic disorder, it is not possible to express $T_c$ in
terms of $\tau_{eff}$ in the {\it whole range} $0\leq T_c\leq T_{c0}$, see
Eq. (\ref{ln(Tc0/Tc)2}). In other words, the combined effect of nonmagnetic
and magnetic scattering on $T_c$ cannot be described by a single universal
parameter depending on the values of $\tau_n, \tau_m^{pot}, \tau_m^{ex}$, and
$\chi$, see Ref. \cite{Openov2} for more details. Hence, while the quantity
$1/\tau_{eff,c}$ characterizes the critical strength of impurity scattering
corresponding to $T_c=0$, the quantity $1/\tau_{eff}$ (when it is less than
$1/\tau_{eff,c}$) doesn't determine the value of $T_c$ unequivocally.

Based on Eqs. (\ref{tau_eff,c}) and (\ref{tau_eff}), it is possible to derive
the following expression for the critical value of $1/\tau_n$ in the presence
of magnetic scattering:
\begin{equation}
\frac{1}{\tau_{n,c}}=\frac{1}{\tau_m^{ex}}\left[%
2\left(\frac{\pi T_{c0}\tau_m^{ex}}{2\gamma}\right)^{1/\chi}-1\right]-%
\frac{1}{\tau_m^{pot}}.
\label{tau_n,c}
\end{equation}
This expression is valid as long as its right-hand side is positive, since
otherwise the superconductivity is completely suppressed solely by magnetic
impurities. The value of $1/\tau_{n,c}$ decreases as $1/\tau_m^{pot}$ and
$1/\tau_m^{ex}$ increase at constant $\chi$ or as $\chi$ increases at
constant $1/\tau_m^{pot}$ and $1/\tau_m^{ex}$.

To conclude this Section, it is interesting to note that $T_c$ doesn't depend
on $\chi$ provided that $1/\tau_m^{ex}=1/\tau_n+1/\tau_m^{pot}$, see
Eq. (\ref{ln(Tc0/Tc)2}).

\vskip 6mm

\centerline{\bf 4. DISCUSSION}

\vskip 2mm

Equation (\ref{ln(Tc0/Tc)2}) is obviously more general than Eqs.
(\ref{ln(Tc0/Tc)3}) and (\ref{ln(Tc0/Tc)4}), which are commonly used for the
analysis of experimental data on $T_c$ suppression by defects and impurities
in HTSCs, see references in the Introduction. In fact, making use of
Eq. (\ref{ln(Tc0/Tc)3}) or Eq. (\ref{ln(Tc0/Tc)4}) one assumes {\it a priori}
that either ($i$) the order parameter in HTSCs is isotropic in momentum
space, or ($ii$) magnetic scatterers are completely absent in HTSCs. In our
opinion, the experimental dependencies of $T_c$ versus impurity concentration
or radiation dose should be analyzed within the framework of the theory
presented above, see Eq. (\ref{ln(Tc0/Tc)2}). One should not {\it guess} as
to the degree of in-plane anisotropy of $\Delta({\bf p})$ and the type of
scatterers, but try to {\it determine} the value of {$\chi$} and relative
weights of magnetic and nonmagnetic components in electron scattering through
comparison of theoretical predictions with available or specially performed
experiments.

We recall that Eq. (\ref{ln(Tc0/Tc)2}) has been derived within the
weak-coupling limit of the BCS model. Note however that the exact solution of
the Eliashberg equations for a particular case of a $d$-wave superconductor
containing nonmagnetic impurities only indicates \cite{Radtke} that the
analytical $T_c/T_{c0}$ versus $1/\tau_n$ curve falls near the numerically
calculated $T_c/T_{c0}$ versus $1/\tau_n^*$ curve, where $1/\tau_n^*$ is the
scattering rate renormalized by the strong-coupling effects (it is
$1/\tau_n^*$ that enters the formula for the experimentally determined
in-plane residual electrical resistivity $\rho_0$). We believe therefore that
Eq. (\ref{ln(Tc0/Tc)2}) is also valid beyond the weak-coupling approximation
implying that $\tau_n$, $\tau_m^{pot}$, and $\tau_m^{ex}$ in
Eq. (\ref{ln(Tc0/Tc)2}) are the {\it renormalized} relaxation times which
govern the experimentally measured physical quantities. It would be
interesting to check this by direct numerical solution of the Eliashberg
equations for an anisotropic superconductor with nonmagnetic and magnetic
impurities.

In order to compare the predictions of theory with experiment, it is
convenient to represent the electron scattering time, Eq. (\ref{tau}), in
terms of the in-plane residual resistivity $\rho_0$. Following Radtke
{\it et al.} \cite{Radtke}, we have
\begin{equation}
\frac{1}{\tau_n}+\frac{1}{\tau_m^{pot}}+\frac{1}{\tau_m^{ex}}=%
\frac{\omega_{pl}^2}{4\pi}\rho_0,
\label{rho_0}
\end{equation}
where $\omega_{pl}$ is the plasma frequency. Note that spin-independent
($1/\tau_n+1/\tau_m^{pot}$) and spin-dependent ($1/\tau_m^{ex}$) scattering
rates variously appear in Eqs. (\ref{ln(Tc0/Tc)2}) and (\ref{rho_0}) for the
critical temperature and residual resistivity. Hence, for a given degree of
anisotropy of the order parameter (i.e., for a given value of $\chi$), the
{\it universal} dependence of $T_c/T_{c0}$ on $\rho_0$ cannot be obtained, as
opposed to the case of a $d$-wave or anisotropic $s$-wave superconductor
containing nonmagnetic impurities only \cite{Radtke,Openov}.

Let us express $\rho_0$ as
\begin{equation}
\rho_0=\rho_0^{nm}+\rho_0^{ex},
\label{rho_0_tot}
\end{equation}
where $\rho_0^{nm}$ is due to electron scattering by nonmagnetic impurities
and potential scattering by magnetic impurities, while $\rho_0^{ex}$ is due
to exchange scattering by magnetic impurities:
\begin{equation}
\frac{1}{\tau_n}+\frac{1}{\tau_m^{pot}}=%
\frac{\omega_{pl}^2}{4\pi}\rho_0^{nm},
\label{rho_0_nm}
\end{equation}
\begin{equation}
\frac{1}{\tau_m^{ex}}=\frac{\omega_{pl}^2}{4\pi}\rho_0^{ex}.
\label{rho_0_ex}
\end{equation}
From Eqs. (\ref{tau_nm}), (\ref{rho_0})-(\ref{rho_0_ex}) we have
\begin{equation}
\rho_0^{nm}=(1-\alpha)\rho_0,~~\rho_0^{ex}=\alpha\rho_0,
\label{rho_0_nm_ex}
\end{equation}
where
\begin{equation}
\alpha=%
\frac{|u_m^{ex}|^2}{(c_n/c_m)|u_n|^2+|u_m^{pot}|^2+|u_m^{ex}|^2}.
\label{alpha}
\end{equation}
The value of $\alpha$ depends, first, on the scattering strengths of
individual nonmagnetic and magnetic impurities (through matrix elements
$u_n, u_m^{pot}, u_m^{ex}$) and, second, on the ratio of impurity
concentrations $c_n/c_m$. The latter is expected to remain constant under
doping or irradiation, at least at relatively low (but sufficient to destroy
the superconductivity) doping level or radiation dose.
For example, low energy irradiation of YBa$_2$Cu$_3$O$_{7-x}$ was found to
induce nonmagnetic defects only \cite{Giapintzakis}, i.e., $c_m/c_n=0$, and
hence $\alpha=0$.

Thus the dependence of $T_c/T_{c0}$ on $\rho_0$ for a given value of $\chi$
is specified by the material-dependent and "disorder-dependent" dimensionless
coefficient $\alpha$. The greater is the relative contribution from exchange
scattering by magnetic impurities to $\rho_0$, the higher is the value of
$\alpha$ ($\alpha$ ranges from 0 in the absence of exchange scattering to 1
in the absence of non-spin-flip scattering). Substituting Eqs.
(\ref{rho_0_nm}) and (\ref{rho_0_ex}) in Eq. (\ref{ln(Tc0/Tc)2}) and taking
Eqs. (\ref{rho_0_nm_ex}) into account, we have
\begin{equation}
\ln\left(\frac{T_{c0}}{T_c}\right)=(1-\chi)\left[\Psi\left(\frac{1}{2}+%
\alpha\frac{\omega_{pl}^2}{8\pi^2 T_c}\rho_0\right)-%
\Psi\left(\frac{1}{2}\right)\right]+\chi\left[\Psi\left(\frac{1}{2}+%
\frac{\omega_{pl}^2}{16\pi^2 T_c}\rho_0\right)-%
\Psi\left(\frac{1}{2}\right)\right].
\label{ln(Tc0/Tc)5}
\end{equation}

Figures 1 - 4 show the plot of $T_c/T_{c0}$ versus $\rho_0$ in a
superconductor with $T_{c0}=100$ K and $\omega_{pl}=1$ eV for different
values of $\chi$ and $\alpha$ ranging from 0 to 1. The choice of $T_{c0}$ and
$\omega_{pl}$ is, to some extent, arbitrary (though these values of $T_{c0}$
and $\omega_{pl}$ are typical for HTSCs, e.g., for YBa$_2$Cu$_3$O$_7$). In
order to go to the other values of $T_{c0}$ and $\omega_{pl}$ one should just
replace $\rho_0$ in Figs. 1 - 4 by $\rho_0(T_{c0}/100)\omega_{pl}^{-2}$,
where $T_{c0}$ is measured in K, and $\omega_{pl}$ is measured in eV.

From Figs. 1 - 4 one can see that at $\chi<1$ the rate of $T_c$ decrease with
increase in $\rho_0$ becomes higher as $\alpha$ increases from 0 to 1, i.e.,
as the relative contribution of exchange scattering to $\rho_0$ increases. At
$\chi=0$ (isotropic $s$-wave pairing) the value of $T_c$ does not depend on
$\rho_0$ for $\alpha=0$, while the superconductivity is completely suppressed
($T_c=0$) at a critical value of $\rho_0^c$ = 1.42 $m\Omega cm$,
113 $\mu\Omega cm$, and 56.5 $\mu\Omega cm$ for $\alpha=$ 0.04, 0.5, and 1
respectively, see Fig. 1. At $\chi=0.5$ (a specific case of anisotropic
$s$-wave or mixed $(d+s)$-wave in-plane pairing) the value of $T_c$
monotonously goes to zero as $\rho_0$ increases for $\alpha=0$, while
$\rho_0^c$ = 401 $\mu\Omega cm$, 113 $\mu\Omega cm$, and 80 $\mu\Omega cm$ for $\alpha=$
0.04, 0.5, and 1 respectively, see Fig. 2. At $\chi=0.8$ (strongly
anisotropic $s$-wave or mixed $(d+s)$-wave in-plane pairing with predominance
of $d$-wave component) one has $\rho_0^c$ = 188 $\mu\Omega cm$,
113 $\mu\Omega cm$, and 99 $\mu\Omega cm$ for $\alpha=$ 0.04, 0.5, and 1
respectively, see Fig. 3. The curves $T_c(\rho_0)$ for different $\alpha$
come closer together as the coefficient $\chi$ increases, i.e., as the order
parameter becomes more anisotropic. At $\chi=1$ ($d$-wave in-plane pairing)
all curves $T_c(\rho_0)$ merge together, see Fig. 4, i.e., the value of
$T_c/T_{c0}$ at a given $\rho_0$ does not depend on $\alpha$, in accordance
with Eq. (\ref{ln(Tc0/Tc)5}), the critical value of $\rho_0^c$ being equal to
113 $\mu\Omega cm$ at any $\alpha$. Note that for $\alpha=0.5$ the curves
$T_c(\rho_0)$ are the same at any value of $\chi$, see Figs. 1 - 4 and
Eq. (\ref{ln(Tc0/Tc)5}).

Magnetic scatterers in a non-$d$-wave superconductor, even if they are
present in a small proportion ($\alpha<<1$), result in $\rho_0^c$ decrease as
compared with $\rho_0^c$ of a sample containing nonmagnetic impurities only.
The decrease in $\rho_0^c$ with $\alpha$ is more pronounced at low values of
$\chi$, i.e., in superconductors having weakly anisotropic order parameter,
see Figs. 1 - 3. At $\chi$ as high as 0.8, i.e., in a superconductor having
strongly anisotropic (but different from a pure $d$-wave) order parameter,
the value of $\rho_0^c$ for $\alpha=1$ is less than twice as low as that for
$\alpha=0.04$, see Fig. 3. In such a superconductor, the role of a small
amount of magnetic impurities is to suppress the superconductivity at a
finite value of $\rho_0^c$ as opposed to the case when exchange scattering is
absent ($\alpha=0$), though the curves  $T_c(\rho_0)$ at $\alpha=$ 0 and 0.04
almost coincide in a very broad range of $T_c/T_{c0}$, see Fig. 3.

In our opinion, an argument in favor of other than pure $d$-wave in-plane
symmetry of the order parameter in HTSCs (at least in some of them) is as
follows. A pure $d$-wave two-dimensional superconductor with $\chi=1$ is
characterized by the {\it universal} dependence of $T_c$ on $\rho_0$ which
is the same at any value of $\alpha$, i.e., at any relative contribution of
exchange scattering to the total value of $\rho_0$, see
Eq. (\ref{ln(Tc0/Tc)5}) and Fig. 4. Meanwhile, $T_c$ versus $\rho_0$ curves
and the values of $\rho_0^c$ in HTSCs are material-dependent and
disorder-dependent \cite{Chien,Walker,Odagawa,Fukuzumi,Brinkmann,%
Rullier,Elesin,Tolpygo,Krash1,Ginsberg,A.G.Sun}. This fact attests that the
value of $\chi$ varies (though, may be, slightly) from one HTSC to another,
while the value of $\alpha$ depends both on the kind of HTSC material and on
the type of impurities or radiation-induced defects.

Besides, the experimentally observed form of $T_c(\rho_0)$ curve in HTSCs is
usually close to linear in a very broad range of critical temperatures
\cite{Odagawa,Fukuzumi,Brinkmann,Rullier,Elesin,Tolpygo,Krash1,Ginsberg,%
A.G.Sun}. The theoretical curve $T_c(\rho_0)$ has such a form if $\chi$ is
close to unity (but $\chi \neq 1$) and $\alpha$ is much less than unity,
e.g., at $\chi=0.8$ and $\alpha=0.04$, see Fig. 3. In contrast, the theory
predicts the negative curvature of $T_c(\rho_0)$ curve for a pure $d$-wave
superconductor (no matter how great is the contribution of exchange
scattering to $\rho_0$), see Fig. 4, as well as for a non-$d$-wave
superconductor with strong exchange scattering, and the positive curvature of
$T_c(\rho_0)$ curve for a non-$d$-wave superconductor containing nonmagnetic
impurities only, see Figs. 1 - 3. So, we expect that the majority of HTSCs
have the mixed $(d+s)$-wave order parameter with predominance of $d$-wave
component ($1-\chi << 1$) and that exchange scattering by magnetic impurities
or radiation-induced defects contributes to $\rho_0$, though quite
insignificantly ($\alpha << 1$).

The admixture of $s$-wave component to a $d$-wave order parameter, e.g.,
$\Delta({\bf p})=\Delta_d (\cos p_x - \cos p_y) + \Delta_s$, may be a
consequence of orthorhombic distortion of CuO$_2$ planes in some HTSCs
\cite{Kouznetsov,Sigrist,Emery}. The value of the coefficient $\chi$ contains
the information about the partial weight of that component in the order
parameter, i.e., about the value of $\Delta_s/\Delta_d$. So, having
determined the value of $\chi$ from experimental data on radiation-induced
and impurity-induced reduction of the critical temperature, one can deduce
the value of $\Delta_s/\Delta_d$ making use of Eq. (\ref{chi}).

Besides, it should be stressed that $(d+s)$-wave symmetry is only one of
possible candidates for the symmetry of anisotropic pairing state in HTSCs.
It is likely to occur in orthorhombic HTSCs. In what concerns purely
tetragonal HTSCs, one may expect mixing of isotropic $s$-wave state with the
state having some higher even angular harmonic, e.g., with $g$-wave state.
Such a mixed $(g+s)$-wave state, just as $(d+s)$-wave state, is also
characterized by $\chi$ values in the range from 0 to 1, depending on the
partial weights of $s$-wave and $g$-wave components in the order parameter.
All the results obtained in this paper are therefore applicable to the case
of $(g+s)$-pairing, as well as to the case of any other in-plane symmetry of
the order parameter.

To conclude this Section, we note that an assumption about the constancy of
the parameter $\alpha$ (i.e., an assumption about the constancy of the ratio
of the concentrations of nonmagnetic to magnetic scatterers) under doping or
irradiation must be checked before detailed comparison of the theory
presented in this paper to experimental data. If this assumption appears to
be incorrect, Eq. (\ref{ln(Tc0/Tc)2}) for the critical temperature can still
be used, the scattering times being given by Eqs. (\ref{rho_0_nm}) and
(\ref{rho_0_ex}). In that case, however, one faces an additional complication
concerning the evaluation of contributions to the residual resistivity
$\rho_0$ from magnetic and nonmagnetic scatterers, $\rho_0^{ex}$ and
$\rho_0^{nm}$ respectively.

\vskip 6mm

\centerline{\bf 5. SUMMARY}

\vskip 2mm

The combined effect of nonmagnetic and magnetic defects and impurities on the
critical temperatures of superconductors with different gap anisotropies was
studied theoretically within the weak coupling limit of the BCS model. For
the case of short-range scattering potentials, an expression was derived
which relates the critical temperature to the relaxation rates of charge
carriers on nonmagnetic and magnetic scatterers as well as to the coefficient
of in-plane anisotropy of the superconducting order parameter on the Fermi
surface.

We note that the results obtained in this paper can be modified to include
the effects of anisotropic (momentum-dependent) impurity scattering. For
example, in the case of significant overlap between the anisotropy functions
of scattering potential and that of the pair potential, the anisotropic
superconductivity has been proven to become less sensitive to nonmagnetic
impurities \cite{Haran1,Kulic,Millis}. However it is not clear if there is
such an overlap in HTSCs.

Besides, numerical calculations within an extended Hubbard model point to the
spatial variation of the order parameter in the vicinity of impurities in
anisotropic superconductors \cite{Franz}. As a result, suppression of $T_c$
is significantly weaker than that predicted by the Abrikosov-Gor'kov-type
theory. This effect presumably is especially pronounced in superconductors
with short coherence length. However, a complete theory of such an effect
remains to be developed.

It is worth noting that impurity doping and irradiation generally result not
only in a structural disorder but also in creation or annihilation of charge
carriers. Thus the effects of carrier and impurity concentrations on $T_c$ of
HTSCs should be considered on equal footing \cite{Loktev}. Moreover, since
high-temperature superconductivity appears upon doping of parent insulators,
a description of those effects should form the basis for the future theory of
HTSCs.

In conclusion, the results obtained provide a basis for evaluation of the
degree of anisotropy of the superconducting order parameter (e.g., for
an estimate of the partial weight of $s$-wave in mixed $(d+s)$-wave order
parameter) as well as the ratio between nonmagnetic and magnetic scattering
rates in high-$T_c$ superconductors through careful comparison of theoretical
predictions with the experiments on impurity-induced and radiation-induced
reduction of the critical temperature. We hope that the present paper will
serve as a stimulus for further experiments on combined effect of nonmagnetic
and magnetic scattering in the copper-oxide superconductors.

\vskip 6mm

\centerline{\bf ACKNOWLEDGMENTS}

\vskip 2mm

This work was supported by the Russian State Program "Integration" and by the
Russian Foundation for Basic Research under Grant No 97-02-16187. The author
would like to thank V. F. Elesin, V. A. Kashurnikov, and A. V. Krasheninnikov
for discussions at the early stage of this work.

\newpage
\centerline{\bf FIGURE CAPTIONS}
\vskip 2mm

Fig. 1. Dependence of the normalized critical temperature $T_c/T_{c0}$ on the
residual resistivity $\rho_0$ due to nonmagnetic and magnetic impurities in a
superconductor with $T_{c0}=100$ K and $\chi=0$ (isotropic $s$-wave pairing)
for different values of the coefficient $\alpha$ specifying the relative
contribution to $\rho_0$ from exchange scattering. $\alpha=0$ (solid curve),
0.04 (long-dashed curve), 0.5 (short-dashed curve); 1 (dot-dashed curve). The
critical value $\rho_0^c$ = 1.42 $m\Omega cm$ for $\alpha=0.04$. The plasma
frequency is taken to be $\omega_{pl}=1$ eV. One can go to the other values
of $T_{c0}$ and $\omega_{pl}$ through replacing $\rho_0$ by
$\rho_0(T_{c0}/100)\omega_{pl}^{-2}$, where $T_{c0}$ is measured in K, and
$\omega_{pl}$ is measured in eV.

Fig. 2. Same as in Fig. 1 for $\chi=0.5$ (a specific case of anisotropic
$s$-wave or $(d+s)$-wave in-plane pairing).

Fig. 3. Same as in Fig. 1 for $\chi=0.8$ (a specific case of anisotropic
$s$-wave or $(d+s)$-wave in-plane pairing).

Fig. 4. Same as in Fig. 1 for $\chi=1$ ($d$-wave in-plane pairing). In this
case the value of $T_c/T_{c0}$ at a given $\rho_0$ does not depend on
$\alpha$, see Eq. (\ref{ln(Tc0/Tc)5}).

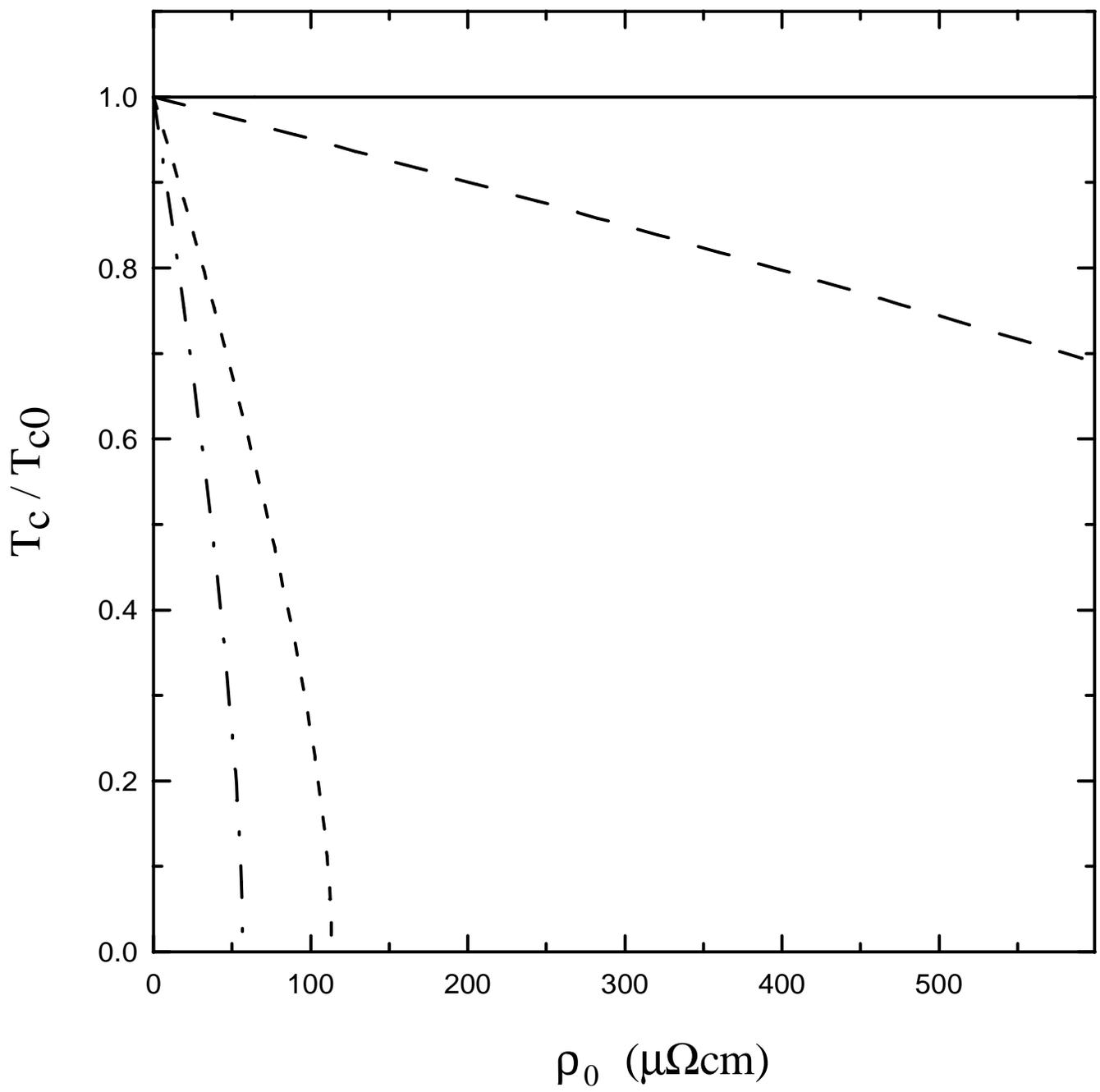
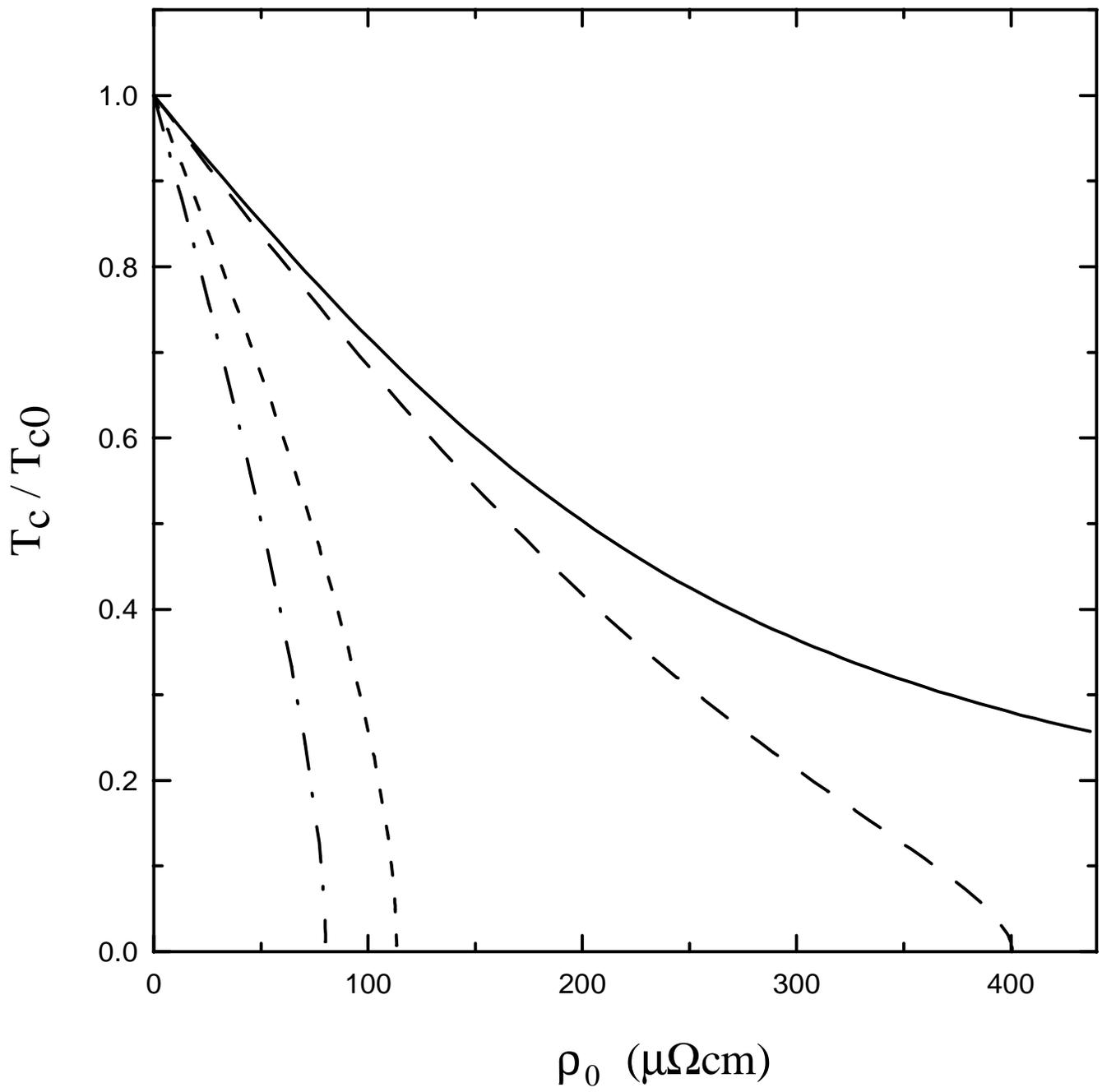
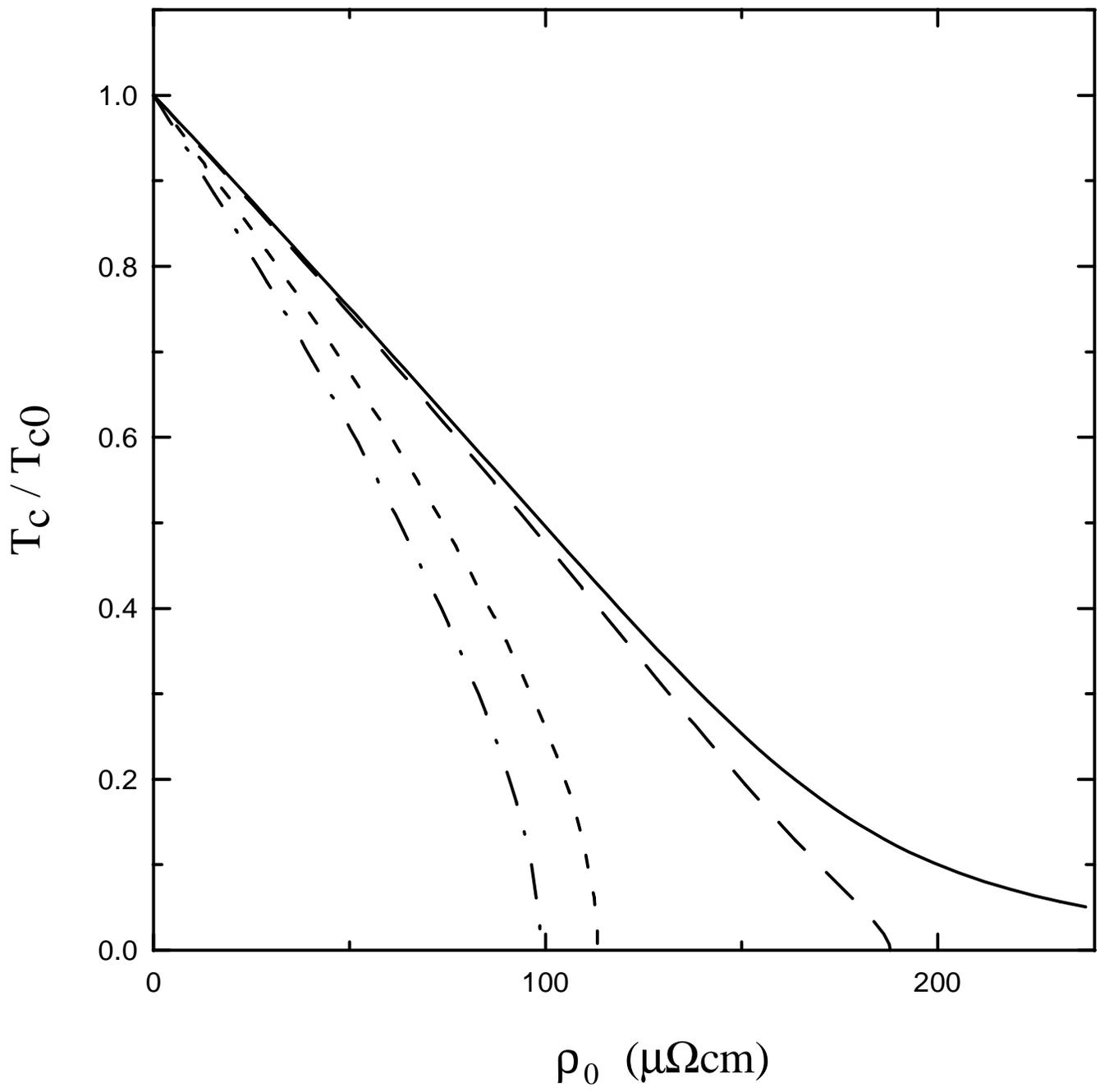
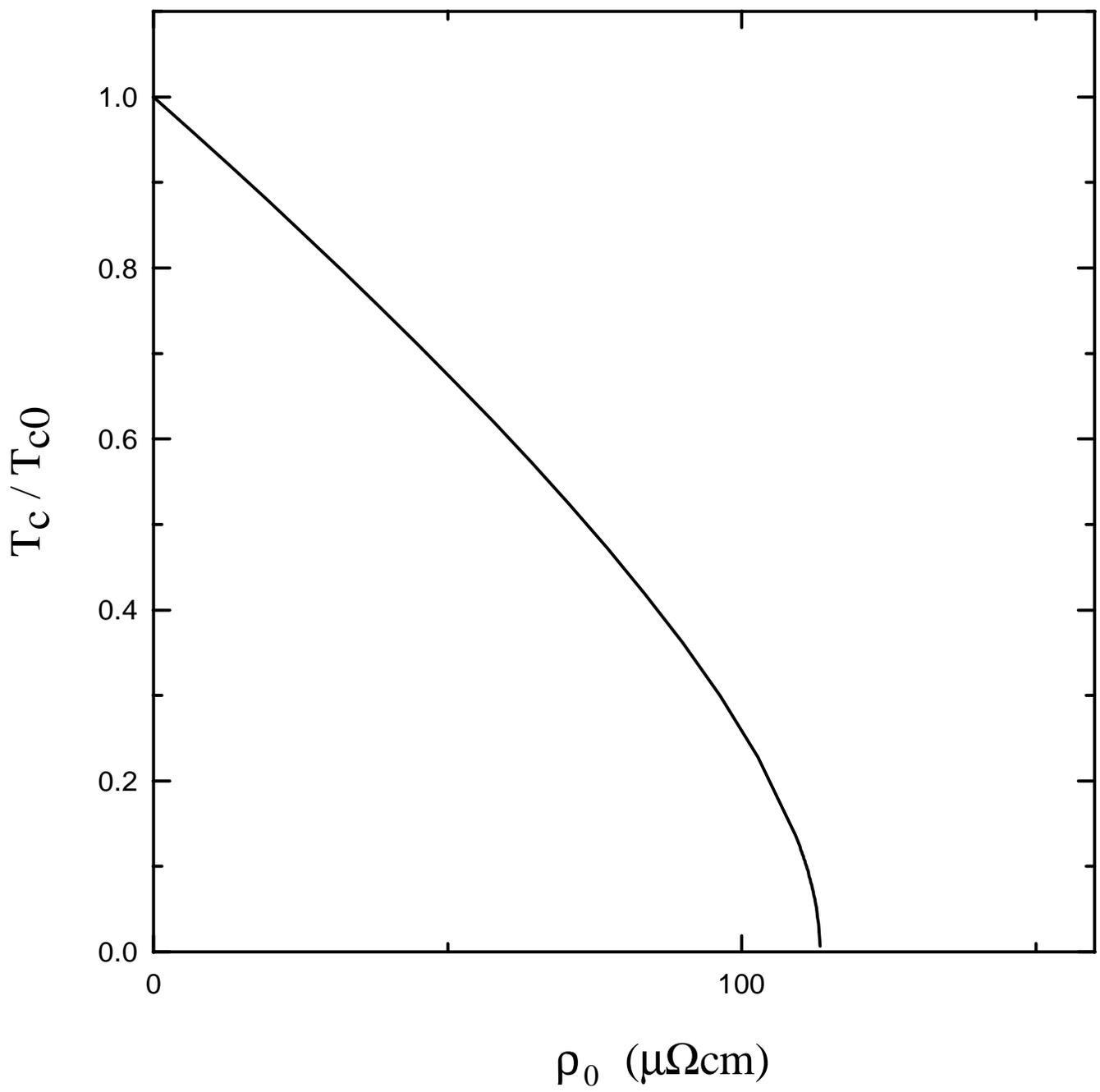
\end{document}